\documentclass[12pt]{article}
\pdfoutput=1

\usepackage{graphicx}
\usepackage{amssymb}
\usepackage{amsmath}
\usepackage{esint}

\usepackage{bm}% bold math

\usepackage{bm}% bold math

\newcommand{\rf}[1]{(\ref{#1})}
\newcommand{\beq}{\begin{equation}}
\newcommand{\beql}[1]{\beq\label{#1}}
\newcommand{\eeq}{\end{equation}}
\newcommand{\bea}{\begin{eqnarray}}
\newcommand{\eea}{\end{eqnarray}}

\newcommand{\ep}{\varepsilon}
\newcommand{\eps}{\epsilon}

\newcommand{\tr}{\mathrm{tr}\,}
\newcommand{\ra}{\rangle}
\newcommand{\la}{\langle}

\newcommand{\cM}{{\cal M}}

\newcommand{\cT}{{\cal T}}

\begin{document}

\begin{center}
\vspace{24pt}
{\large \bf The microscopic structure of 2D CDT  coupled to matter}
\vspace{30pt}

{\sl J. Ambj\o rn}$\,^{a,b}$,
{\sl A. G\"{o}rlich}$\,^{a,c}$,
{\sl J. Jurkiewicz}$\,^c$ and
{\sl H. Zhang}$\,^{c}$

\vspace{48pt}

$^a$~The Niels Bohr Institute, Copenhagen University\\
Blegdamsvej 17, DK-2100 Copenhagen \O , Denmark.\\
email: ambjorn@nbi.dk, goerlich@nbi.dk\\

\vspace{10pt}

$^b$~Institute for Mathematics, Astrophysics and Particle Physics (IMAPP)\\
Radbaud University Nijmegen, Heyendaalseweg 135,
6525 AJ, \\
Nijmegen, The Netherlands

\vspace{10pt}

$^c$~Institute of Physics, Jagiellonian University,\\
Reymonta 4, PL 30-059 Krakow, Poland.\\
email: jerzy.jurkiewicz@uj.edu.pl, zhang@th.if.uj.edu.pl\\

\vspace{96pt}
\end{center}

%\addtolength{\baselineskip}{0.20\baselineskip}
%\vspace{2cm}

\begin{center}
{\bf Abstract}
\end{center}

\noindent

We show that for   1+1 dimensional  Causal Dynamical 
Triangulations (CDT) coupled to 4 massive scalar fields one can 
construct an effective transfer matrix if the  masses squared is 
larger than or equal to  0.05. The properties of this transfer matrix
can explain why CDT coupled to matter can behave completely different 
from ``pure'' CDT. We identify the important critical 
exponent in the effective action, which may determine the universality 
class of the model.

\newpage

\section{Introduction}

Causal Dynamical Triangulations (CDT) in 1+1 dimensions 
can be considered a toy model for more advanced models of quantum gravity. 
The simplest version of the model,  pure fluctuating geometry without matter,
can be solved analytically\footnote{There are a number of 
generalizations, also only dealing with fluctuating geometries, which
can be solved analytically \cite{generalizations}.} \cite{ambjorn}. 
The random geometries  present in the CDT path integral 
are constructed by gluing together flat simplices (triangles) in 
such a way that one has  a global time foliation. 
The topology of space at a given time is assumed to be $S^1$.
This topology is preserved in the time evolution. In all our 
considerations time is Wick rotated and the triangles used are 
assumed to be equilateral with edge lengths $a$. We label time 
with integer numbers $t$. The geometry of a spatial slice at time $t$  
is completely characterized by its length, i.e. in the CDT model 
by $n(t)$ - the number of edges forming the spatial $S^1$ . 
In the path integral representation of the time evolution, 
spatial states at integer times $t$ and $t+1$ 
are connected in all possible ways consistent with the foliation. 
In the case of CDT without matter fields the time evolution is 
generated by a transfer matrix 
$\langle n_t | \cM | n_{t+1}\rangle = \exp(-L(n_t,n_{t+1}))$ 
with correctly normalized spatial states $|n\rangle$\footnote{The 
normalization should include the symmetry factor of the states,
in the original paper they were realized as marking one of the 
triangles in $\langle n(t)|$. It is possible to take a symmetric definition, 
which we do in this paper. In the following we assume 
the norm $\langle n | m \rangle = \delta_{nm}$.}. 
The explicit expression for $L(n,m)$ can be 
interpreted as a term of the effective action, since in obtaining 
it we sum over all geometrical realizations joining the two states.

One may view the same geometry using a dual trivalent lattice, 
where  vertices are located in the centers of triangles and links 
being dual to the edges in the original lattice. 
Each vertex has exactly three neighbors, two
at the same time $t$ (which can be considered as a half-integer time 
with respect to the original triangulation) and one at time $t\pm 1$. 
As before the links at the same time value form a closed spatial 
geometry $S^1$, the quantities $n_t$ and $n_{t+1}$ represent  
numbers of links pointing up or down from the line at $t + 1/2$. 
The dual formulation is completely equivalent to the original one
and we will use it in this article.

The Hilbert-Einstein action for a triangulation $\cT$, $S(\cT)$, provides 
the weight $\exp(-S(\cT))$  to be assigned to the triangulation
in the path integral. In  1+1 dimensions  there is no curvature
term (it is a topological invariant), and we are left with the 
cosmological term, which on the lattice takes to form
\begin{equation}
S(\cT) = \sum_i \lambda n_i = \lambda N, \quad n_i \equiv n(t_i),
\end{equation}
where $N$ is the  total number of triangles (dual vertices) 
of the triangulation $\cT$. $\lambda$  is the  dimensionless 
bare cosmological constant. The weight contains a factor 
\begin{equation}
e^{-\frac{\lambda}{2}(n+m)} = g^{n+m},~~~g \equiv e^{-\frac{\lambda}{2}},
\end{equation} 
in each transfer matrix element $\langle n| \cM | m\rangle$. This 
factor compensates the entropy factor, always present in 
the models of quantum geometry, where the number of 
different triangulations between $n$ and $m$ typically
(for large $n$ and $m$)
grows as $\exp(\lambda_c N)$ with some critical $\lambda_c$. 
The quantum amplitude (the partition function) can now be written as
\begin{equation}
Z = \sum_{\cT} \exp(-S(\cT)) = \exp\left(-\sum_t L(n_t,n_{t+1})\right)
\end{equation}

In \cite{zhang-2}-\cite{zhang-4}) we studied the effect of coupling
$d$  massive scalar fields to the 1+1 dimensional CDT model. 
We assumed the fields to be located at the  vertices of the dual lattice
and introduced the partition function  
\begin{equation}
Z = \sum_{\cT}  \int \prod_{i,\mu} d\phi_i^{\mu}
\;\exp\left(-\lambda N_{\cT} - S_{matter}(\phi_i^{\mu})\right)
\label{partition}
\end{equation}
with the Gaussian  matter action 
\begin{equation}
S_{matter} = \sum_{l_{ij},\mu} (\phi_i^{\mu}-\phi_j^{\mu})^2 + 
m^2 \sum_{i,\mu} \left(\phi_i^{\mu}\right)^2,
\label{action}
\end{equation}
where $l_{ij}$ denote the links in the dual lattice and $\mu=1,\ldots,d$.

It is clear that the integral over the fields (\ref{partition}) 
is well defined for $m^2 > 0$.
In order to see the role of the mass parameter in the action $S_{matter}$, 
we redefine the field variables by $\psi_i^{\mu} = m \phi_i^{\mu}$, 
and the  matter action changes to
\begin{equation}
S_{matter} = \frac{1}{m^2} 
\sum_{l_{ij},\mu} (\psi_i^{\mu}-\psi_j^{\mu})^2 + \sum_{i,\mu} 
\left(\psi_i^{\mu}\right)^2
\label{action1}
\end{equation}
From (\ref{action1}) it follows that the mass controls the 
range of  field interactions: in the large mass limit 
$m^2 \rightarrow \infty$, the couplings between neighboring vertices 
can be neglected, and as a consequence, 
the contribution of matter can be eliminated ($d=0$, pure gravity), 
while in the small mass limit $m^2 \rightarrow 0$ (the massless case), 
the range of interaction becomes long.
It is also clear that there may no longer exist a transfer matrix
depending only on the geometric variables $n_t$ at time slice $t$ and 
its neighboring time slices $t\pm 1$. Integrating out the matter 
degrees of freedom might introduce long range interactions between
various time slices, an effect which one would expect to increase
with decreasing the mass. Indeed, the effect of small or zero 
mass matter fields on the 1+1 dimensional global geometry is dramatic
when $d>1$ as reported in \cite{zhang-2} (and also seen in earlier 
studies using other kind of matter fields \cite{aal}). One observed
the appearance of a ``semi-classical'' de Sitter-like blob 
with Hausdorff dimension $d_H=3$, much like what has been observed
in higher dimensional CDT \cite{higher-blob} (see \cite{review} for a review).
Surprisingly, even for this system there seems to exist an 
{\it effective} local action which couples only neighboring 
time slices and which describes the ``semi-classical'' blob
and the fluctuations around it.

Inspired by these results we will try clarify to which extent we for 
$d >0$ can talk about an effective transfer matrix depending only on
the geometric variables $n_t$, and 
try to understand which  characteristic features of the transfer 
matrix change when a  semiclassical blob is created for  $d > 1$.

More precisely we will study 1+1 dimensional CDT with $d=4$ and $m^2=0.05$.
This  mass is so small that a blob is formed and still so  large that  
an effective transfer matrix depending only on $n_t$ can relatively 
easily be extracted (for smaller masses this becomes increasingly difficult).
We will compare our results to that of 1+1 dimensional CDT without
matter fields where no blob is formed.

The form of the transfer matrix for pure 1+1 dimensional CDT 
will be an indication of the form to be expected for the other cases.
It is \cite{ambjorn,review}
 \begin{equation}\label{jan0}
 \langle n | \cM | m\rangle = \binom{n+m}{n}\sqrt{\frac{4 n m}{(n+m)^2}}
\;g^{n+m} = e^{-L(n,m)}.
 \end{equation}
We are interested in the asymptotic limit, when $n,~m$ are large 
but $(n-m)$ stays finite. We obtain
\begin{equation}
 L(n,m) = C - (n+m)\log (2g) + \frac{1}{2}\frac{(n-m)^2}{n+m} + V(n+m)
 \label{pure}
 \end{equation}
 where the potential
\begin{equation}
V(n+m) =  \frac{1}{2}\log(n+m) +\frac{1}{4(n+m)}+O(\frac{1}{(n+m)^3})
 \label{puregravity}
\end{equation}
In (\ref{pure}) we see a term $- (n+m)\log (2g) = -(n+m)\log(g/g_c)$, 
a linear term related to  the entropy of states. 
A similar entropy term will always be present and 
we will fine tune the value of $g$ to be
as close to $g_c$ as possible in numerical analysis 
aimed at the determination of the transfer matrix elements.
The next term is a ``kinetic'' term, coupling the spatial 
volumes at slices $t$ and $t+1$. This term is ``local'' 
and we also expect such a term to be present in the effective 
transfer matrix. 
Finally we have a ``potential'', diagonal term $V(n+m)$. 
The leading large volume  term in $V$ is a logarithmic term. 
The rest of the terms decrease for large spatial volume.
One can show  that if we observe a blob 
with $d_H=3$ one cannot have terms of the type $(n+m)^\alpha$ 
with $0 < \alpha < 1$ in $V$. As a consequence 
the value of the parameter
in front of the potential $\log(n+m)$ is important for the  
global behavior of the model, as we will explicitly show later. 
The small-volume corrections may be important for a detailed behavior 
of the system at small volumes. We will not be concerned with the 
detailed analytic form of these corrections for CDT coupled to matter, but  
will determine them numerically by Monte Carlo simulations. More precisely
we parameterize the transfer matrix 
\beq\label{jan1}
\langle n | \cM^{th} | m \rangle = \exp(-L_{eff}(n,m))
\eeq 
as
\begin{equation}
 L_{eff} (n, m) = C-(n+m) \log (g/g_c)+
\mu \log(n+m)+ \frac{1}{\Gamma} \frac{ (n-m)^2}{(n+m)} ,
\label{tm-effective}
\end{equation} 
for $(n+m) > K$ 
for some  $K$, while for $(n+m) \leq K$ the transfer matrix 
is simply determined from the computer simulations. 
In an overlap region we match the assumed, parameterized transfer matrix
to the numerically determined one and in this way we determine the 
coefficients $\mu$ and $\Gamma$, which will depend on the number 
of Gaussian fields, $d$, and the value of the mass squared, $m^2$.
Once the constants in $L_{eff}$ are determined we can use the 
transfer matrix \rf{jan1} for arbitrary large $n+m >K$.  

If a transfer matrix $\la n|\cM | m\ra$ exists also for systems
with matter coupled to the geometry then the partition function 
for a system with periodic boundary conditions in the time variable 
with period $T$ is given by
\begin{equation}
Z(g, T)=\sum_{\{n_i\}} \langle n_1 | \cM | n_2\rangle
\langle n_2 | \cM | n_3 \rangle\cdots \langle n_T | \cM | n_1\rangle 
= \tr \cM^T, \quad n_i \equiv n(t_i)
\label{jan2}
\end{equation}
This function is defined for $g < g_c$ and 
the limit $g \to g_c$ corresponds to taking a large volume limit.
Essentially, $Z(g,T)$ is the partition function we will use 
when checking that the transfer matrix model produces the 
same spatial volume distributions as the full model defined by 
\rf{partition}.

The rest of this article is organized as follows.
In the next Section we discuss the numerical methods used to obtain 
the estimate for $g_c$ and to determine the elements of the transfer 
matrix. To see if the 
transfer matrix determined this way reproduces the
observed distributions of spatial volumes we perform 
Monte Carlo volume experiments, 
using the determined transfer matrix for systems with a fixed volume $N$ 
and periodicity $T$, comparable to that used
in the paper \cite{zhang-2}.
In  Section 3 we analyze the eigenvalue spectrum of the 
transfer matrix for $g \to g_c$ for pure CDT and for CDT with 4 
scalar massive field with mass $m^2=0.05$. We recall the exact dependence
of the spectrum for pure gravity and compare it with the $d=4$ case. 
We summarize the results in section 4.

\section{Determination of the transfer matrix}

As described above we assume that after integrating out the matter fields in 
\rf{partition} that the quantum geometry can be described by an effective 
transfer matrix $\cM$ with  matrix elements $\langle n | \cM | m \rangle$. 
The matrix elements are (semi-)positive and the matrix is symmetric. 
The last property follows from the invariance of the model 
under the change of  time arrow. We will try  to determine 
the effective transfer matrix from Monte Carlo simulations. 
We use the numerical set-up  described in \cite{zhang-2,zhang-3,zhang-4}.
For a periodic system with the period $T$
the probability to observe a sequence of spatial volumes 
$\{n_1,n_2,\dots n_T\}$ is
\begin{equation}
P_T(n_1,n_2,\dots n_T) = 
\frac{\langle n_1 | \cM | n_2\rangle\langle n_2 | \cM | n_3 \rangle 
\cdots \langle n_T | \cM | n_1 \rangle}{\tr \cM^T}
\label{transfer-calculate}
\end{equation}
This quantity is invariant under cyclic permutations and reversion of 
order\\ $n_1,n_2,\ldots,n_T\to n_T,\ldots,n_2,n_1$. 
The probability to observe a particular combination 
of spatial volumes can be measured by Monte Carlo simulations.
Using such measured probabilities for a sequence of periods $T$ 
we can determine $\langle n | \cM | m \rangle$ up to a
multiplicative constant. We compare two measured probabilities
\begin{equation}
P_{T,1}(n,m) = 
\frac{\langle n | \cM | m \rangle\langle m | \cM^T | n \rangle}{\tr \cM^{T+1}}
\end{equation}
where we determine the probability to observe volumes $n$ and $m$ 
in two neighboring slices for a system with the period $T+1$ and
\begin{equation}
P_{2T,T}(n,m) =  \frac{\langle n | \cM^T | m \rangle\langle m | \cM^T | n \rangle}{\tr \cM^{2T}}
\end{equation}
where we determine the probability to observe volumes $n$ and $m$ in two slices separated by $T$ steps for a system with the period $2T$ .
We determine the transfer matrix elements from
\begin{equation}
\label{jan3}
\langle n | \cM | m \rangle  = C \frac{P_{T,1}(n,m)}{\sqrt{P_{2T,T}(n,m)}}
\end{equation}
where $C$ is the multiplicative factor mentioned above. 
Notice that the lhs of \rf{jan3} is independent of $T$.

\begin{figure}[h]
\begin{center}
\includegraphics[width=0.43\textwidth]{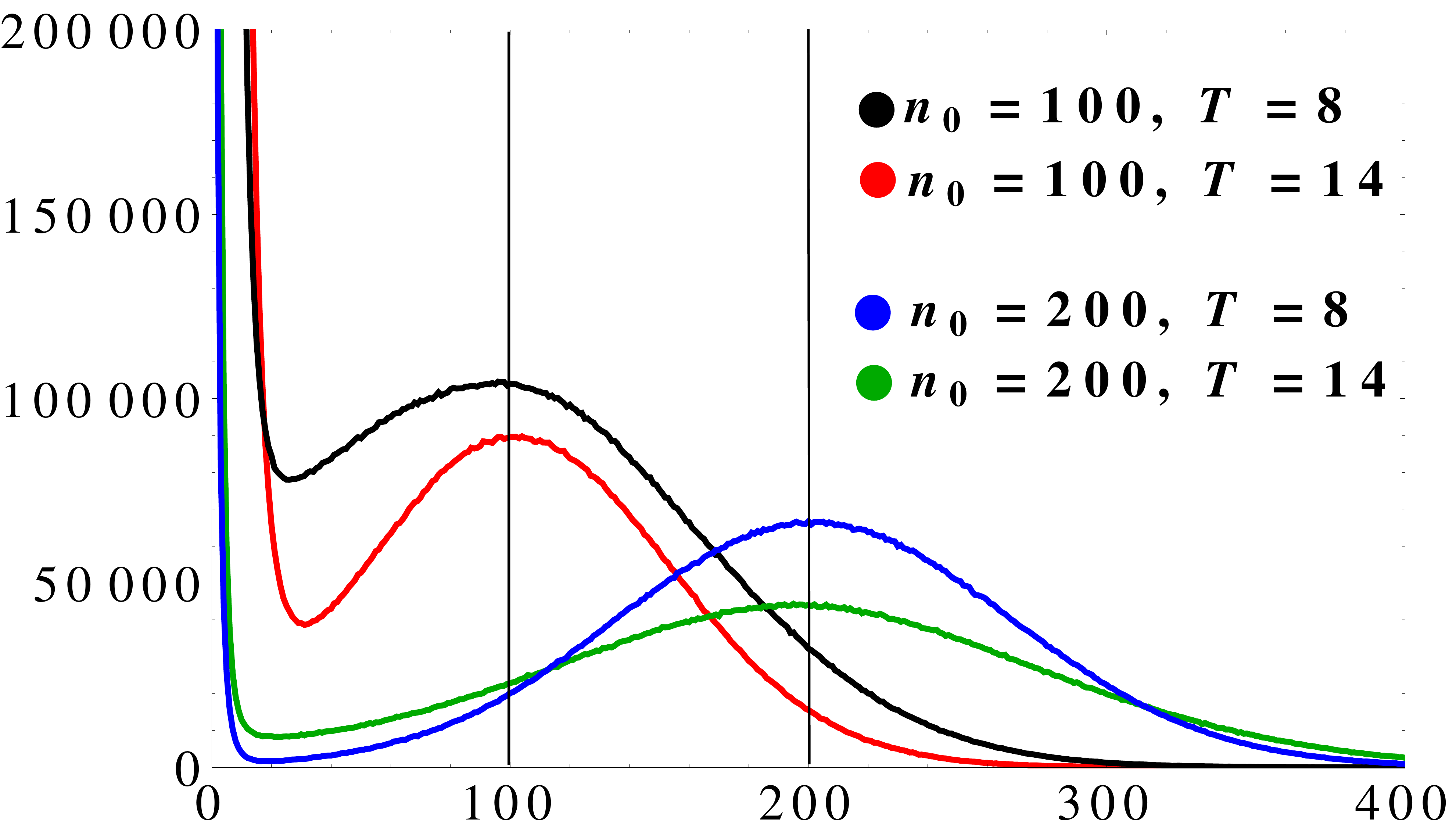}
\end{center}
\caption{Averaged volume distribution in a slice for a 
system with $d=4$ and $m^2=0.05$. We take $n_0=100, 200$ 
and check the location of a maximum using $T=8$ and $T = 14$.}
\label{element}
\end{figure}

The transfer matrix depends on the parameter $g$ and we expect the behavior 
$\langle n | \cM | m \rangle \propto (g/g_c)^{n+m}$. We are interested 
in the  for $g \to g_c$, where
the average volume diverges (the continuum limit). To avoid the problem
of infinite volume we introduce in the simulations  a modified action
\begin{equation}
S(\cT) \to  S(\cT) + \ep \sum_{t=1}^{T} (n_t -n_0)^2.
\label{fix}
\end{equation}
This modification forces the spatial volume at each slice to 
fluctuate around $n_0$. 
The new action leads to a modified transfer matrix but one can 
easily recover the original transfer matrix,  as will be explained below.
The modification of the action serves two purposes. 
Firstly, it allows us to determine the critical coupling $g_c$.
The average volume distribution in the slices, for $T$ and $n_0$ sufficiently
large,  should have a maximum at $\langle n\rangle = n_0$. 
For finite $T$ we observe different distributions for
different choices of $n_0$, but for $g=g_c$ the maximum should 
be stable. Typical distributions are shown in Fig.\ \ref{element} for
a system with four massive scalar fields with a mass $m^2=0.05$ and $g = g_c$.
This method can be used to determine the critical parameter $g_c$ 
even in cases, where the transfer matrix may not be a good approximation. 
In the table \ref{compare-g} we list the estimates for a range of 
mass parameters $m^2$ and, for comparison, for pure gravity.
In the following we use  $g=g_c$.
\begin{table}
	\begin{center}
		\begin{tabular}{|c|c|c|}
\hline
$m^2$ 	&$g_c$ estimates  &  the best estimate $g_c$\\ \hline 
0.00       & 0.2988-0.3000& 0.2991 \\ \hline
0.05       &0.3205-0.3212& 0.3210 \\ \hline
0.10       &0.33375-0.3338& 0.33378 \\ \hline
0.15       & 0.34412-0.34417& 0.34415 \\ \hline
0.20       & 0.35284-0.35285 & 0.35285 \\ \hline
5.00       & 0.48322-.048342 & 0.48341 \\ \hline
Pure gravity&  0.4999-0.50 & 0.50\\ \hline
\end{tabular}
\end{center}
\caption{Estimated values of $g_c$. The estimates come from volume distributions with $n_0= 0,~ 50,~ 100, ~150,~ 200$ and $T = 20$.}
\label{compare-g}
\end{table}

Secondly, the modification \rf{fix} makes the spatial volume fluctuations
much more controllable in the Monte Carlo simulations. 
It changes the probability distributions to 
\begin{equation}
\tilde{P}_T(n_1, n_2, \dots, n_T)
= \frac{\langle n_1 | \tilde{\cM} | n_2 \rangle \langle n_2 |
\tilde{\cM} | n_3 \rangle \cdots
\langle n_T | \tilde{\cM} | n_1 \rangle}{\tr \tilde{\cM}^T},
	\label{EqTildePTM}
\end{equation}
where the relation between $\cM$ and $\tilde{\cM}$ is given by
\begin{equation}\label{ja22}
	\langle n | \cM | m \rangle =
C' e^{\frac{1}{2}\epsilon (n - n_0)^2}
\langle n | \tilde{\cM} | m \rangle e^{\frac{1}{2}\epsilon (m - n_0)^2},
\end{equation}
again up to an arbitrary factor $C'$. 
This relation permits us to determine the matrix elements 
$\langle n | \cM | m \rangle$. The method will work 
for $n,~m$ restricted to a window around $n_0$, 
where volume fluctuations are small.
The size of the window depends on the parameter $\epsilon$ 
and typically outside this window the fluctuations become large. 
In practice we measure the matrix elements for a sequence of
$n_0$ and connect the results by requiring the best overlap. 

\begin{figure}[h]
\begin{center}
\includegraphics[width=0.45\textwidth]{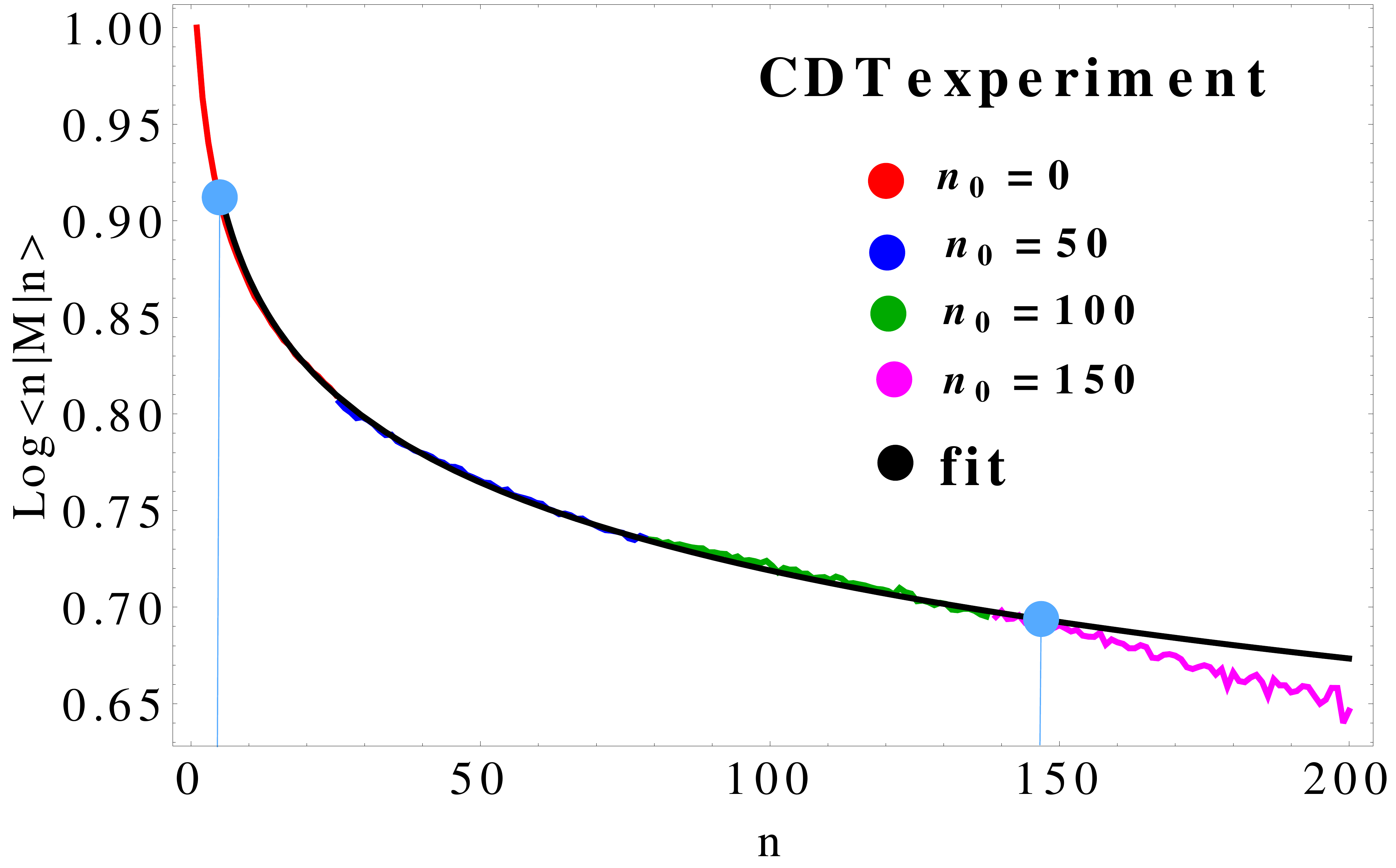}
\includegraphics[width=0.43\textwidth]{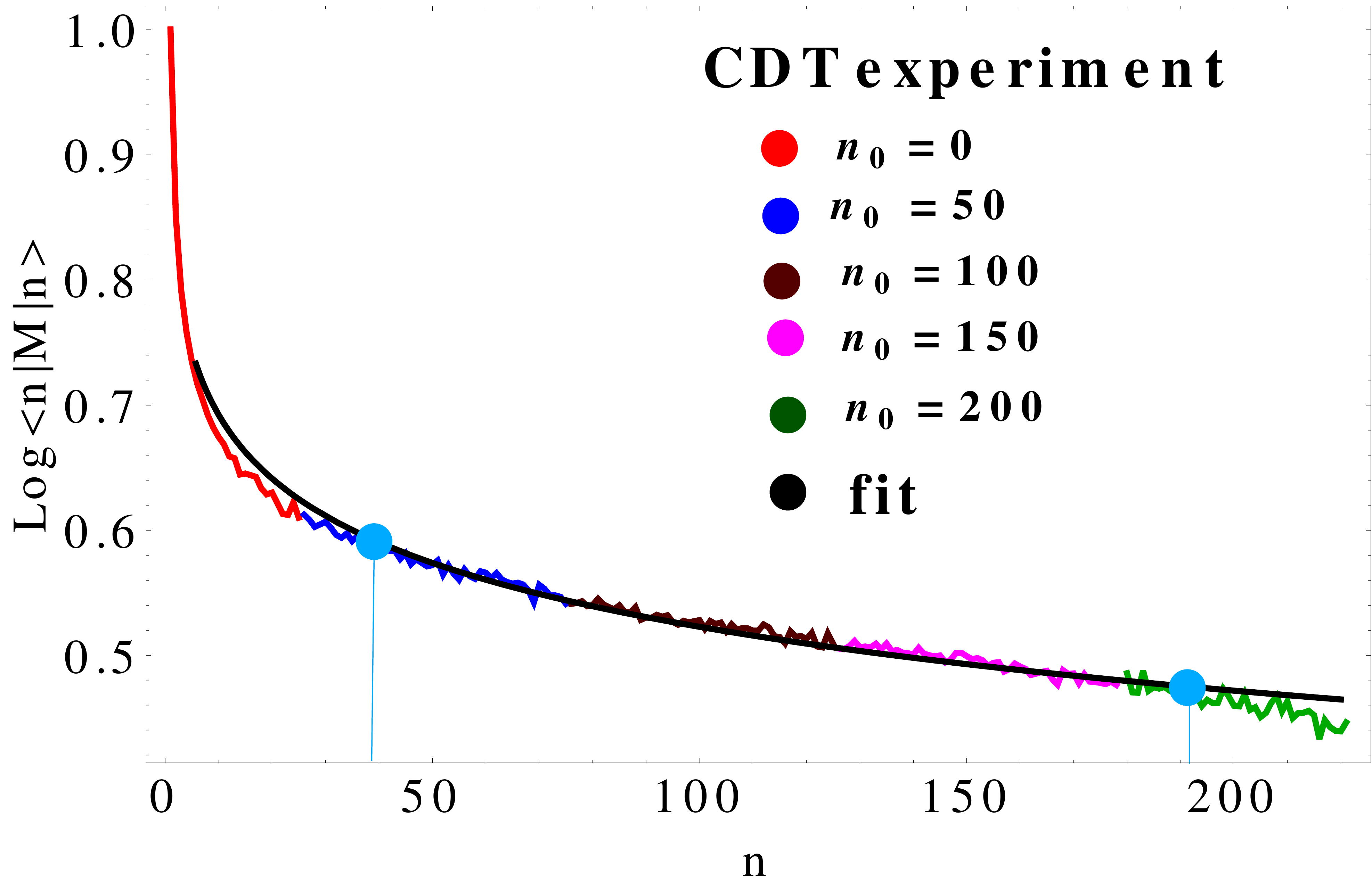}
\end{center}
\caption{The logarithm of the diagonal of the transfer matrix at $g=g_c$ 
for pure gravity (left) and for $d=4$, $m^2=0.05$ (right). 
The plots show a combination of results obtained with  
$n_0= 0, ~50, ~100, ~150$. The black line represents a fit
to the asymptotic power behavior of the diagonal part, 
with the coefficient determined in the range limit by the blue dots.}
\label{tm-scaled}
\end{figure}

The  transfer matrix, \rf{jan1}-\rf{tm-effective}, 
depends on the two parameters $\mu$ and $\Gamma$. 
We determine $\mu$ by measuring the diagonal part of the 
transfer matrix (i.e. $\langle n | \cM | n \rangle$), where the 
kinetic term vanishes. We use $T=8$.
For $T >5$ the results seem insensitive to the periodic boundary
conditions imposed in the time direction.
As before we combine  the measurements for 
various overlapping windows around different values of $n_0$.
The results of best fits are shown in Fig.\  \ref{tm-scaled} for pure gravity 
(as a test of the method) and for $d=4$,  $m^2=0.05$.

\begin{figure}[h]
\begin{center}
\includegraphics[width=0.4\textwidth]{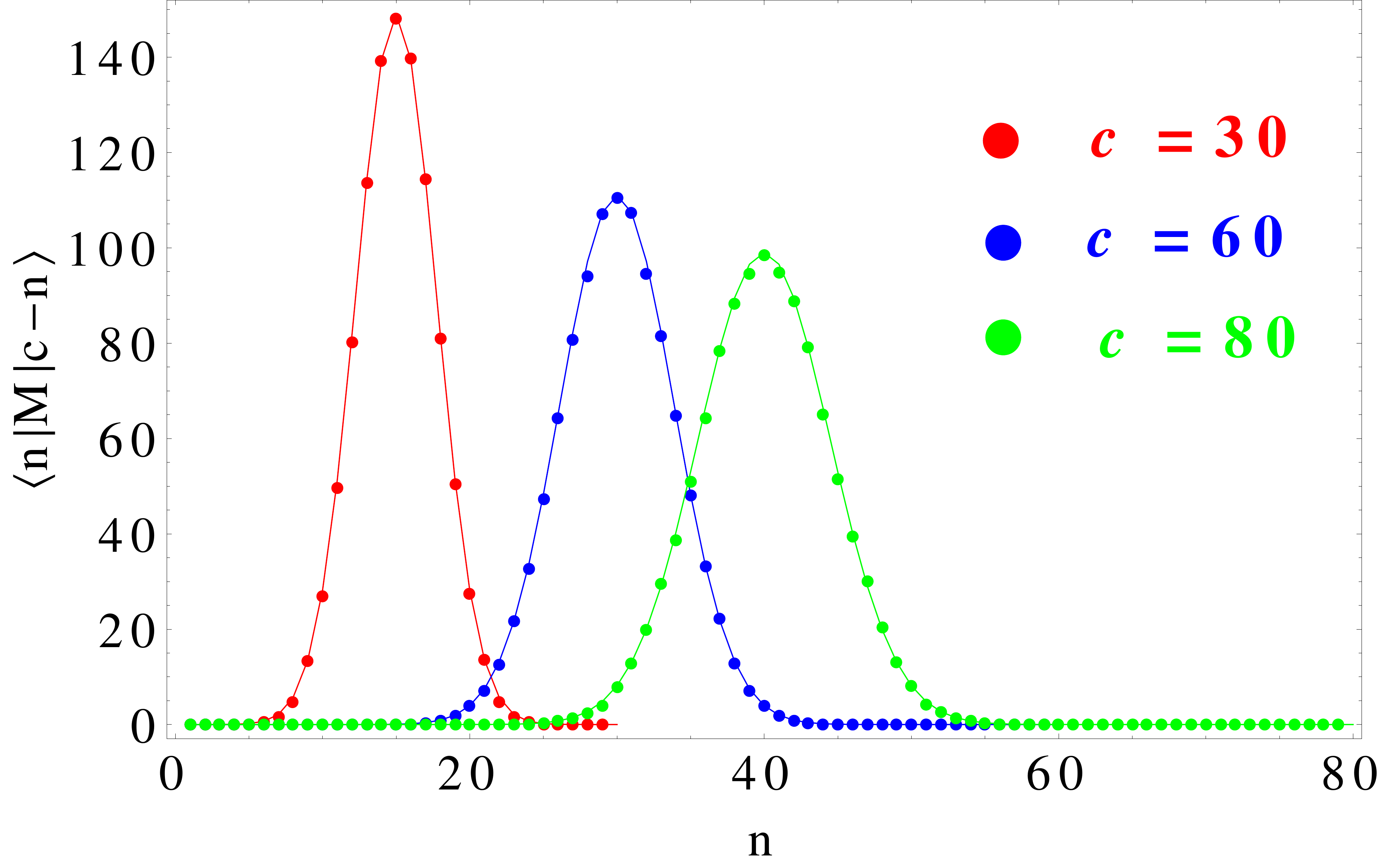}
\includegraphics[width=0.4\textwidth]{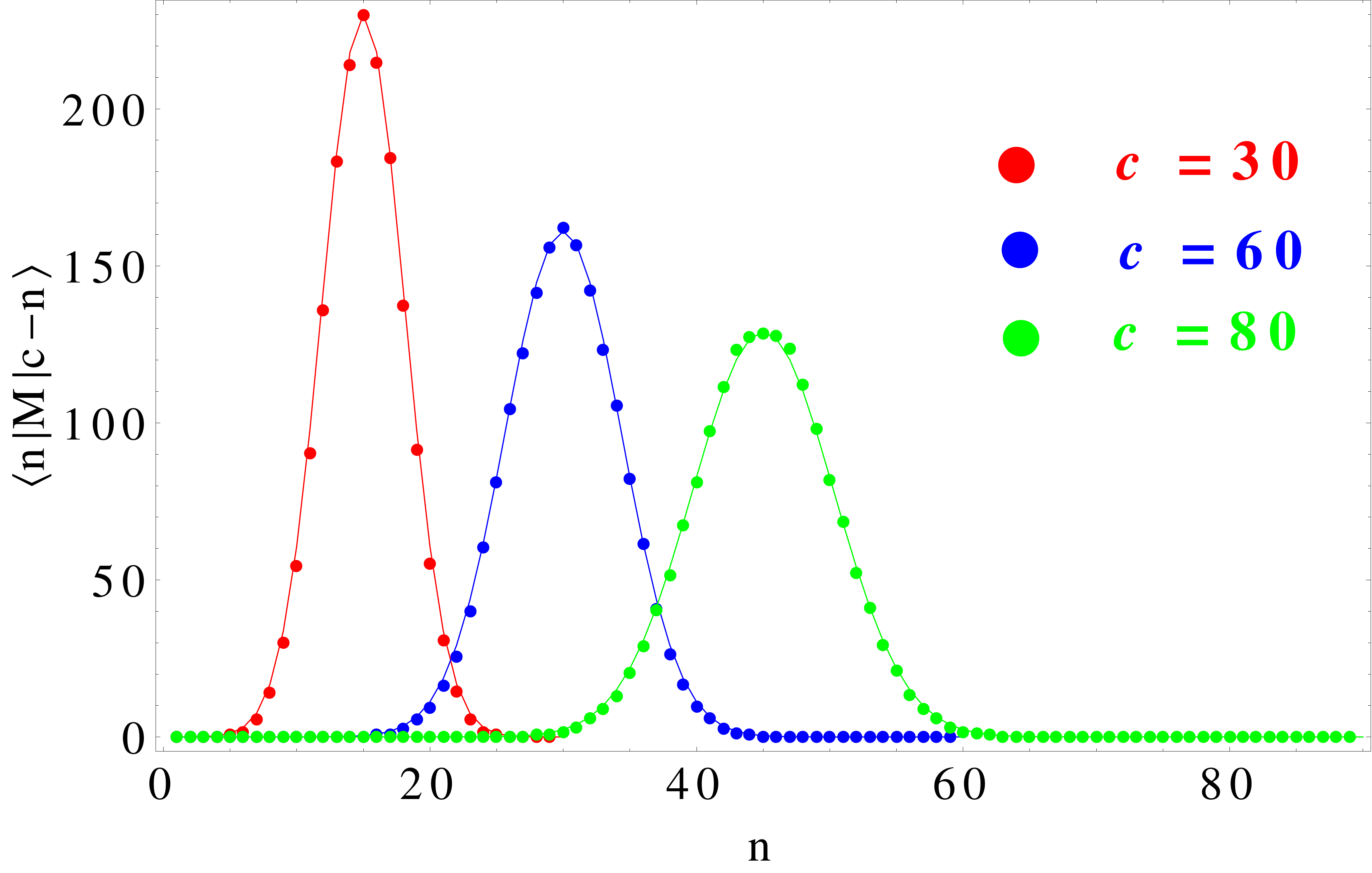}
\end{center}
\caption{The transfer matrix $\langle n | M | c - n \rangle$ 
for various $c's$: pure gravity (left) and $d=4$, $m^2 =0.05$ (right).}
\label{off-fitting-range}
\end{figure}

\begin{figure}[h]
\begin{center}
\includegraphics[width=0.4\textwidth]{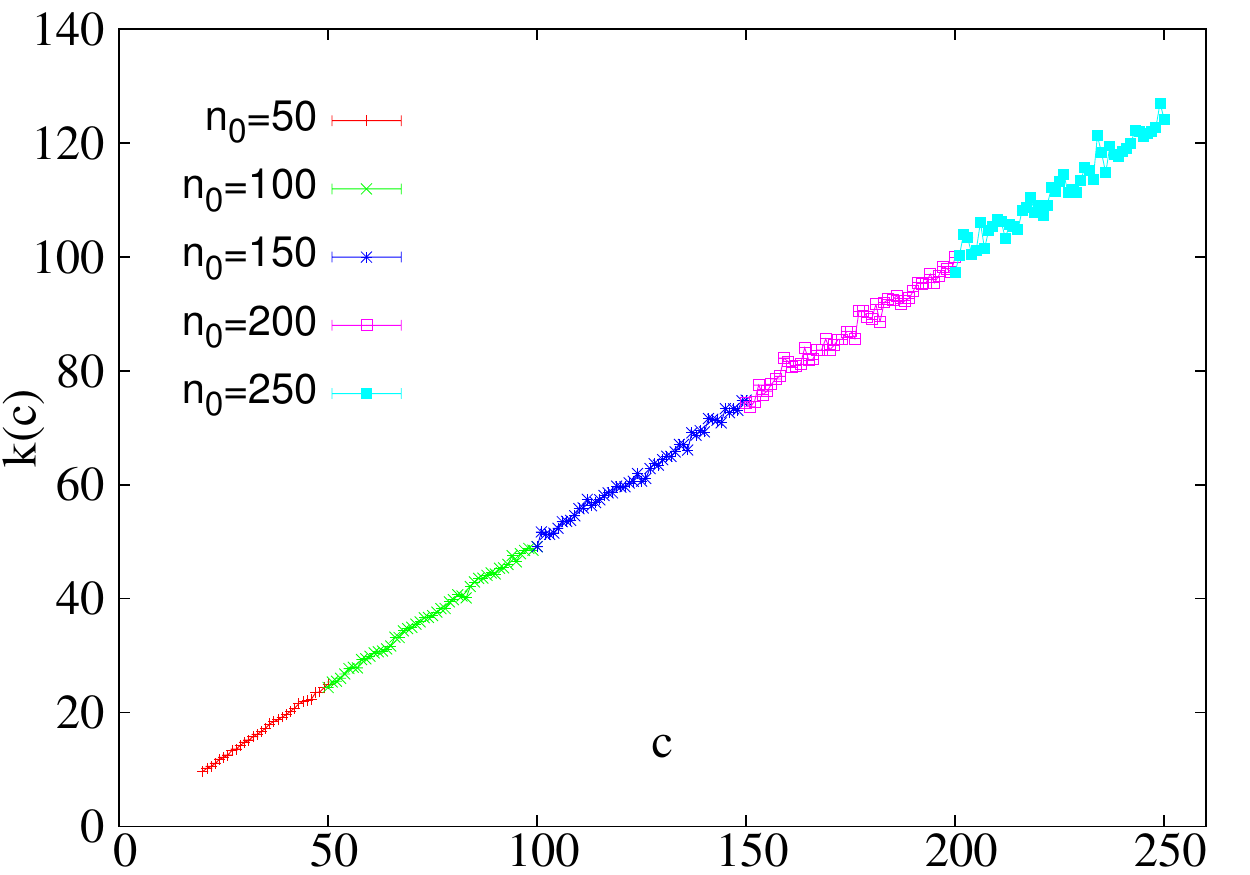}
\includegraphics[width=0.4\textwidth]{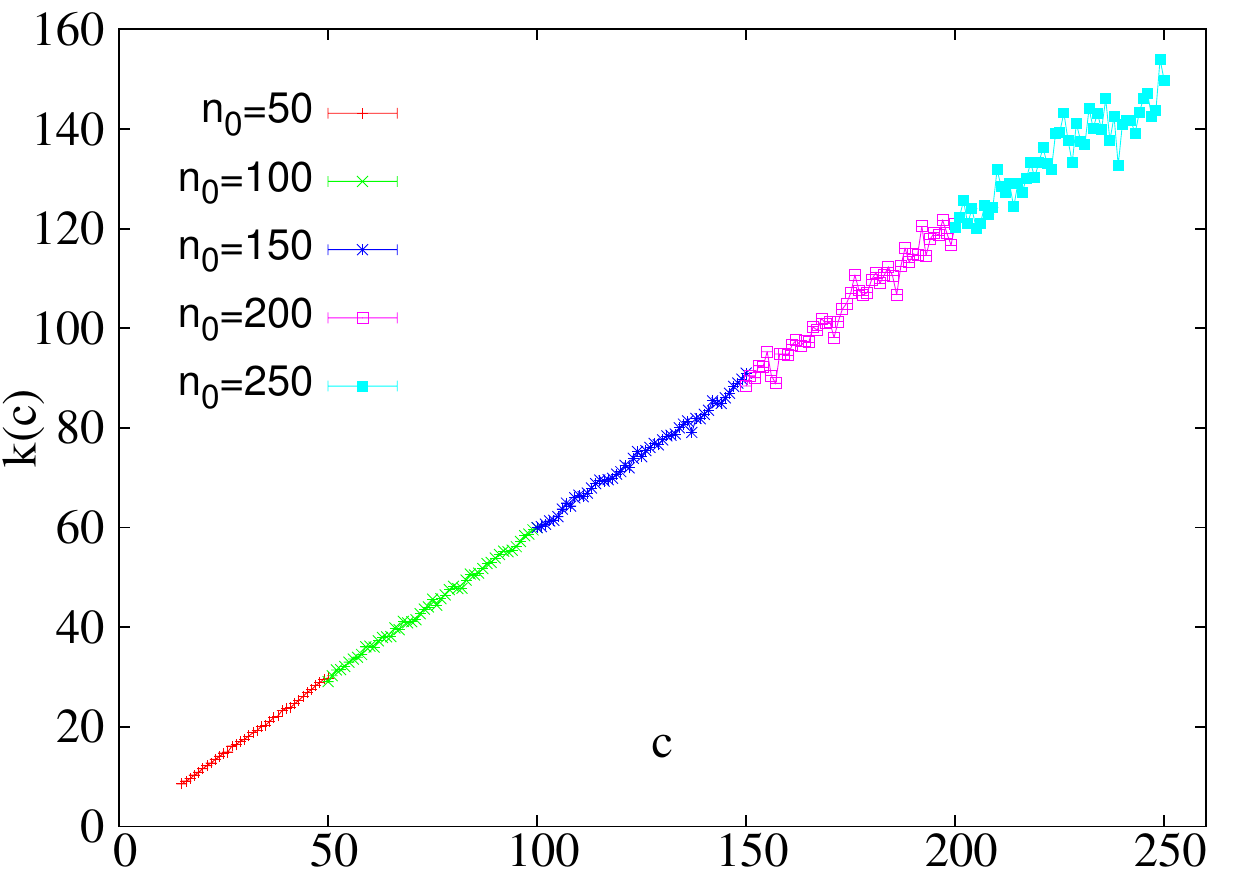}
\end{center}
\caption{Linear behavior of $k(c) \approx \Gamma c$ for pure gravity (left) 
and $d=4$, $m^2 =0.05$ (right).}
\label{linear}
\end{figure}

We determine $\Gamma$ by measuring the transfer matrix 
for $n+m$ fixed (such that only the kinetic term changes).
For sufficiently large $c$ we have 
\beql{gauss}
\langle n | M | c - n \rangle =
\mathcal{N}(c) \exp \left[{- \frac{(2n-c)^2}{ \Gamma c}}\right], 
\eeq
where the terms in the effective action which only depend on $c$ 
are included in the normalization. 
This is illustrated on  Fig.\  \ref{off-fitting-range}, 
and we observe that the width of the Gaussian grows with $c$. 
This growth is with a very high accuracy linear, 
as is illustrated on the figure \ref{linear}.
The fitted values of the parameters $\mu$ and $\Gamma$ are 
listed in the table \ref{parameters}.
\begin{table}
\begin{center}
\begin{tabular}{|c|c|c|c|c|c|c|c|c|}
\hline
$m^2$   & 0.05 & 0.10 & 0.15 & 0.20 & 5.0  & Pure Gravity \\ \hline
$\mu$  & 0.64628 & 0.59365 &  0.55432  &  0.53233 & 0.503512  & 0.505763   \\ \hline
$\Gamma$  & 2.5 $\pm 0.05$ & 2.2  $\pm 0.1$ & 2.17  $\pm 0.15$ & 2.15  $\pm 0.1$ & 2  $\pm 0.1$ & 2 $\pm 0.05$  \\ \hline
\end{tabular}
\end{center}
\caption{Estimated values of the parameters $\mu$ and $\Gamma$ as 
explained in the text.}
\label{parameters}
\end{table}

\begin{figure}[h]
\begin{center}
\includegraphics[width=0.4\textwidth]{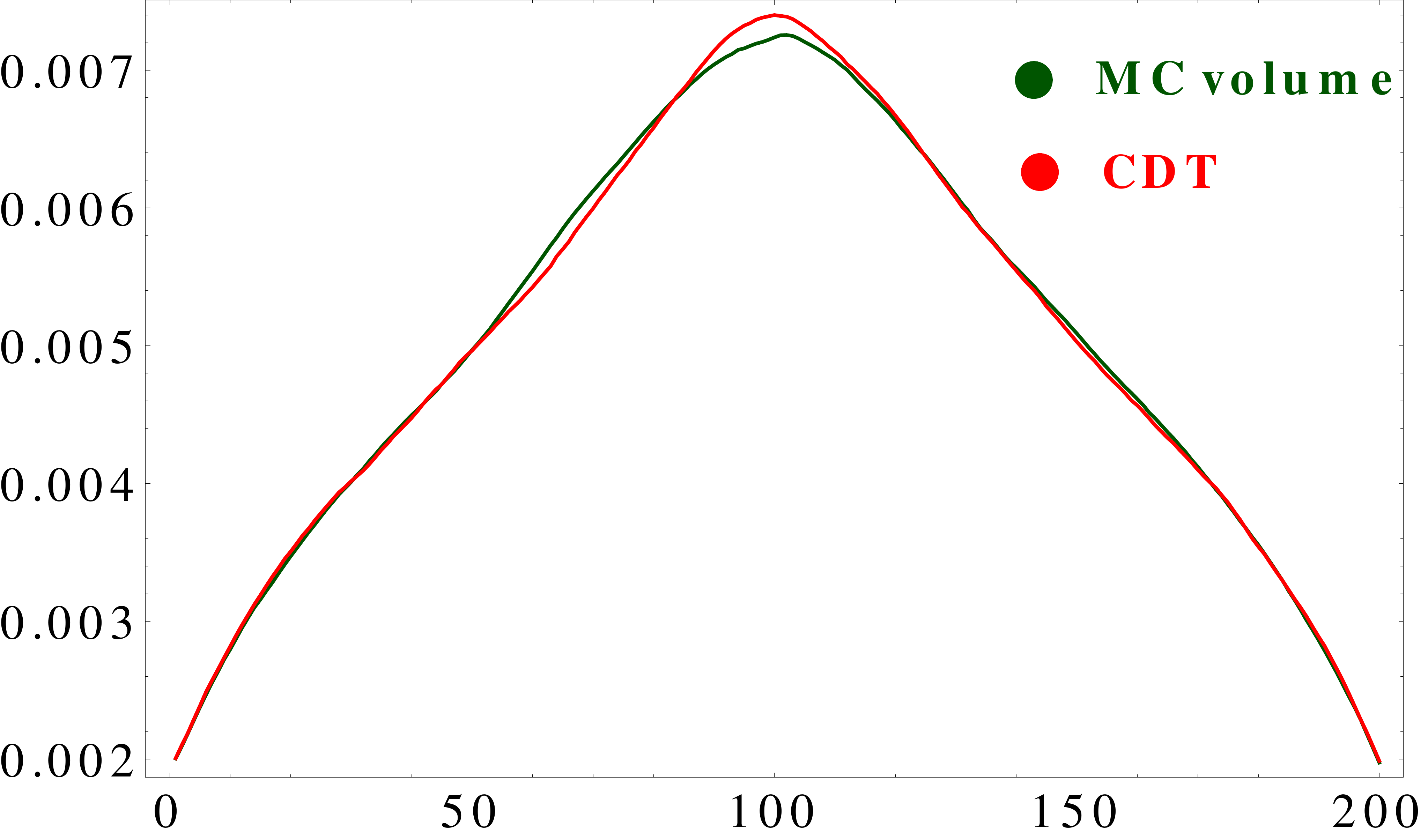}
\includegraphics[width=0.4\textwidth]{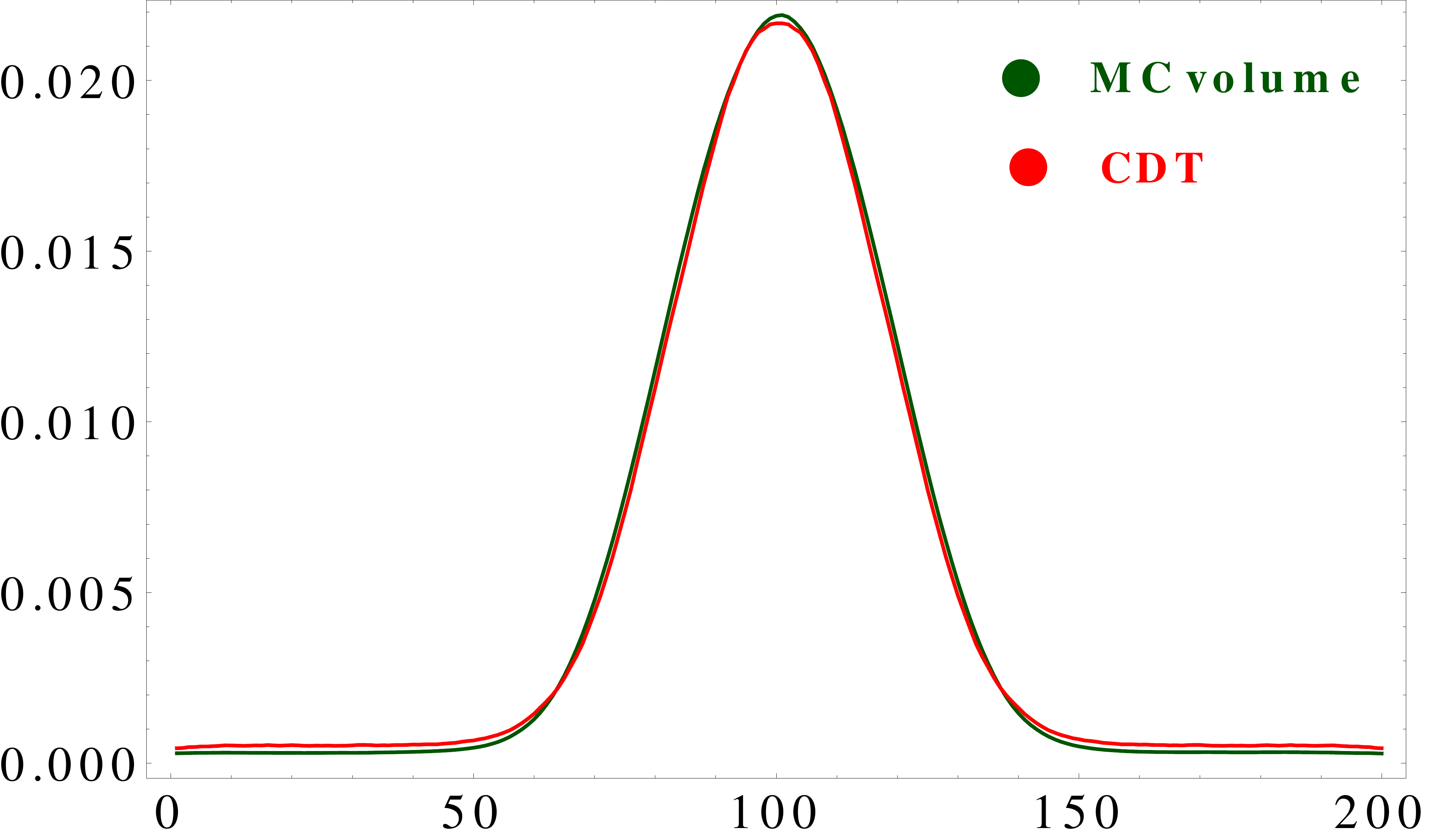}
\end{center}
\caption{Comparison of $n(t)/N$ using the effective transfer matrix
and full CDT:  pure gravity (left) and $d=4$, $m^2=0.05$ (right). 
We use $N=16000$, T=200 for both cases.}
\label{topology}
\end{figure}

As a check of the quality of the transfer matrix determined this 
way we performed a Monte Carlo simulation where
the probability  assigned to a geometry with  
spatial volumes $\{n_1,n_2,\dots,n_T\}$ is 
\begin{equation}
P(n_1,n_2,\dots,n_T) \propto 
\langle n_1 | \cM^{eff} | n_2\rangle  
\langle n_2 | \cM^{eff} | n_3\rangle\cdots  
\langle n_T | \cM^{eff} | n_1\rangle e^{-\eps (\sum_t n_t -N)^2}.
\label{MCvolume}
\end{equation}
The ``effective'' transfer matrix entries 
$ \langle n_1 | \cM^{eff} | n_2\rangle$ used in \rf{MCvolume} are 
\begin{eqnarray}
\label{reconstructed}
\langle n | \cM^{eff} \ m \rangle =
\begin{cases}
\langle n |\cM^{exp} | m\rangle ,      & m+n \leq K \\
\langle n |\cM^{th} | m \rangle  ,     &  m+n >K  .
\end{cases}
\end{eqnarray}
The small volume part is obtained numerically   
and the large volume part is given by 
\rf{jan1} and \rf{tm-effective} with $\mu$ and $\Gamma$ determined
as described above. $K$ is chosen in the range 50-100. 
The extra factor in \rf{MCvolume} is added to enforce the total volume to 
fluctuate around a given value $N$. The resulting
distribution can be compared to the one obtained using 
the full partition function \rf{partition}. 
We use here  the same method to analyze
spatial volume profiles as that presented in \cite{zhang-2,zhang-3,zhang-4}: 
each configuration is shifted in such a way that the ``center of mass''  
of the spatial volume distribution is placed at time $t=T/2$.
In this way we produce an artificial maximum also for pure gravity
where there is no blob, 
but the properties of this distribution is quite different from
the ``real'' blob distribution, as explained in \cite{zhang-2,zhang-3,zhang-4}
and as  seen from Fig.\ \ref{topology}.
The agreement is very good. 
For the case $d=4$, $m^2=0.05$ the small volume part of the transfer 
matrix is important in order to get a quantitative agreement of
the volume profile in the tail of the distribution 
(the agreement  becomes less good at a quantitative 
level if $K<50$ in \rf{reconstructed}) 
The existence or non-existence of the blob is however 
entirely linked to the value of $\mu$ in the effective 
action \rf{tm-effective}: simply using this effective 
action (with $g=g_c$) and adding a small volume term $1/(n+m)$ 
like in \rf{puregravity} in order to stabilize the logarithmic
term for $n+m=0$ one obtains by Monte Carlo simulations a transition 
from a non-blob phase to a blob phase simply by changing $\mu$.
$\mu$ is like a critical exponent: it is well know that 
$\mu =1/2$ appears as an entropy in CDT \cite{ambjorn}, and this is
seen explicitly from \rf{pure} for large $n$:
\beq\label{jan4}   
\la n|\cM |n\ra \sim (2n)^{-1/2} (g/g_c)^{2n}(1+O(1/n)).
\eeq
Changing the details of the triangulations used in CDT, changing the 
scale of $n$ etc, will change $g_c$ but not the exponent $\mu$.
Similarly, when we couple CDT to the matter fields we have 
\beq\label{jan5}   
\la n|\cM |n\ra \sim (2n)^{-\mu} (g/g_c)^{2n}(1+O(1/n)).
\eeq
where $\mu$ will invariant under changes of triangulation
details, but will depend on $d$ and $m$. Our numerical results
suggest  that there is a critical value  $\mu_c$ such that for $\mu > \mu_c$ 
the geometry change universality class from that of pure CDT
(with Hausdorff dimension $d_H =2$) to that the ``blob'' geometry
(which has $d_H=3$).

\section{Eigenvalue spectrum  of the transfer matrix}

\begin{figure}[h]
\begin{center}
\includegraphics[width=0.4\textwidth]{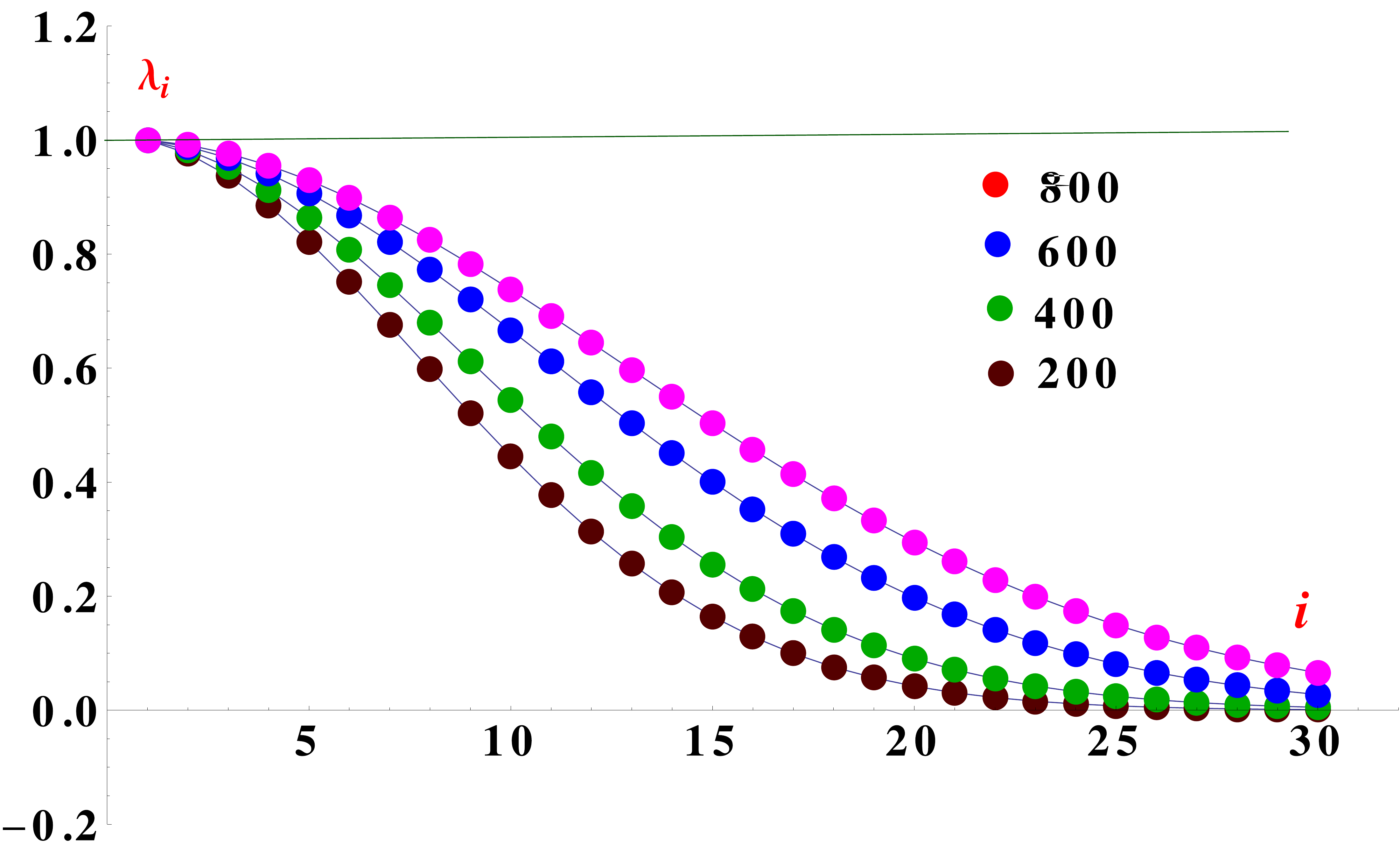}
\includegraphics[width=0.4\textwidth]{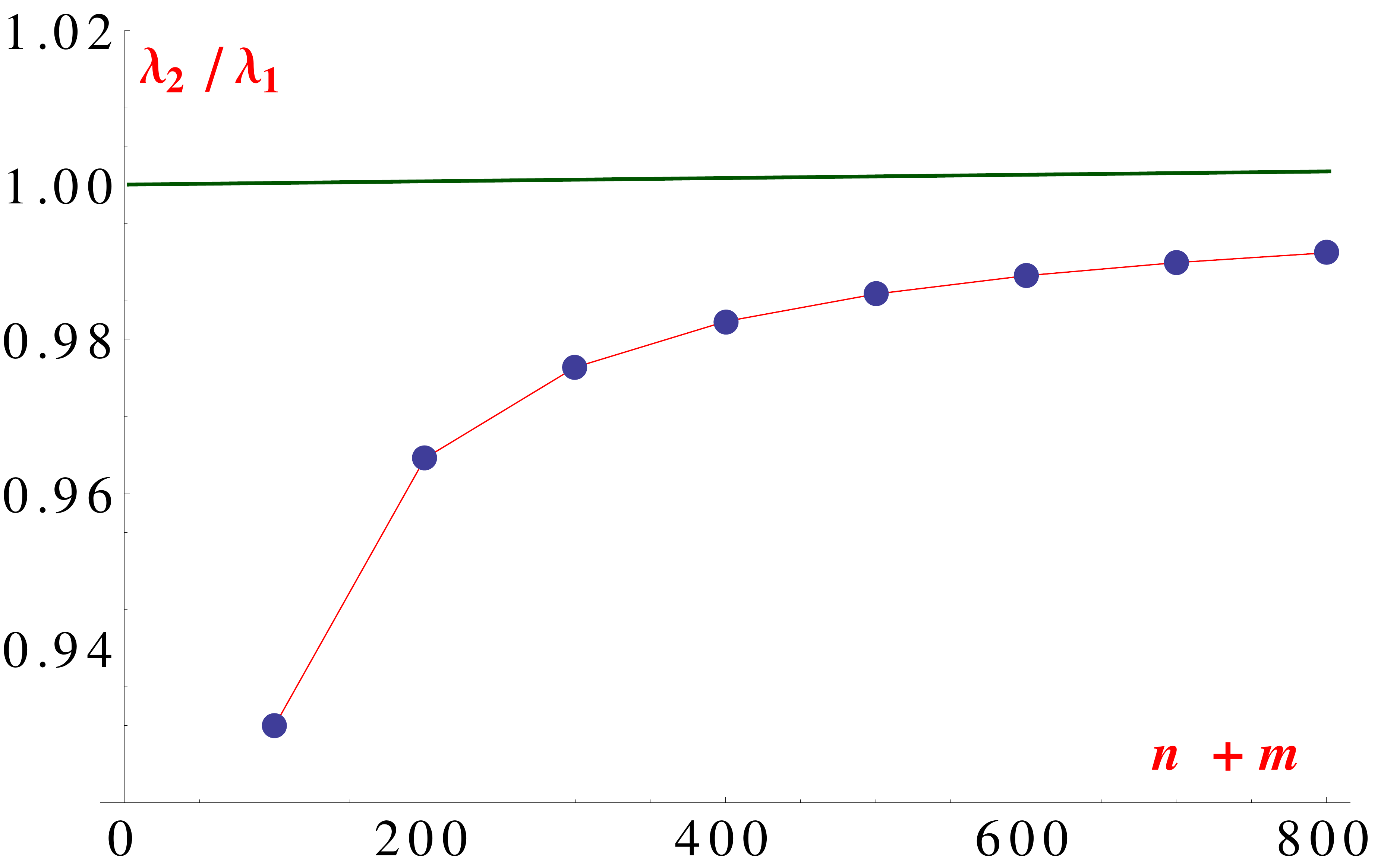}
\end{center}
\caption{Left: Eigenvalues for pure gravity at $g=g_c$ with cut-off $n_{max}= 200, 400, 600, 800$. Right:  $\lambda_{2}/\lambda_{1}$ as a function of $n_{max}$ .}
\label{eigenfunction-pure}
\end{figure}

The analytic solution of  1+1 dimensional  CDT \cite{ambjorn} permits us 
to determine the eigenvalue spectrum
of the exact transfer matrix as a function of $g$. Using the parametrization
\begin{equation}
g = \frac{1}{2 \cosh(\beta)},\quad g_c= \frac{1}{2}
\end{equation} 
and solving the eigenvalue equation we find (see also \cite{charlotte} for 
the eigenvalue spectrum for a more general model) 
\begin{equation}
\lambda_n = e^{-(2n+2)\beta},\quad n=1,2,\dots
\end{equation}
The effective transfer matrix determined above for $d=4$
was only obtained up to a normalization so we can only determine
the ratio of eigenvalues  for this ``empirical'' transfer matrix.
For the pure CDT model we have $\lambda_n/\lambda_1 = \exp(-2n\beta)$.
Thus the ratio goes to zero exponentially with $n$ for $g<g_c$. 
For $g=g_c$ the spectrum becomes degenerate.
In a numerical analysis we will never be able to achieve this limit, 
since our numerical transfer matrix is necessarily finite and 
the dependence on $n_{max}$ for $g=g_c$
for pure gravity is illustrated on the figure \ref{eigenfunction-pure}.
We can see that at the critical point the convergence of the 
ratios to one is slow as function of $n_{max}$. 
%For $g<g_c$ the situation is different and numerically we may 
%obtain values $\lambda_n/\lambda_1$ as a function of $g$ quite
%close to the critical point. The estimates systematically get 
%worse for larger $n$, but as can be seen on Fig.
%\ref{lambda-pure} the dependence of the estimates on $g$ follows
%very well the theoretical prediction for small $n$.
%\begin{figure}[h]
%\begin{center}
%\includegraphics[width=0.45\textwidth]{eigen-fitting-pure.pdf}
%\end{center}
%\caption{ The dependence of $\lambda_{2}/\lambda_{1}$ and  $\lambda_{3}/\lambda_{1}$ on $g$ for pure gravity. The points correspond to the extrapolated transfer matrix (see eq. (\ref{reconstructed})) and the
%lines come from the theory.}
%\label{lambda-pure}
%\end{figure}

\begin{figure}[h]
\begin{center}
\includegraphics[width=0.5\textwidth]{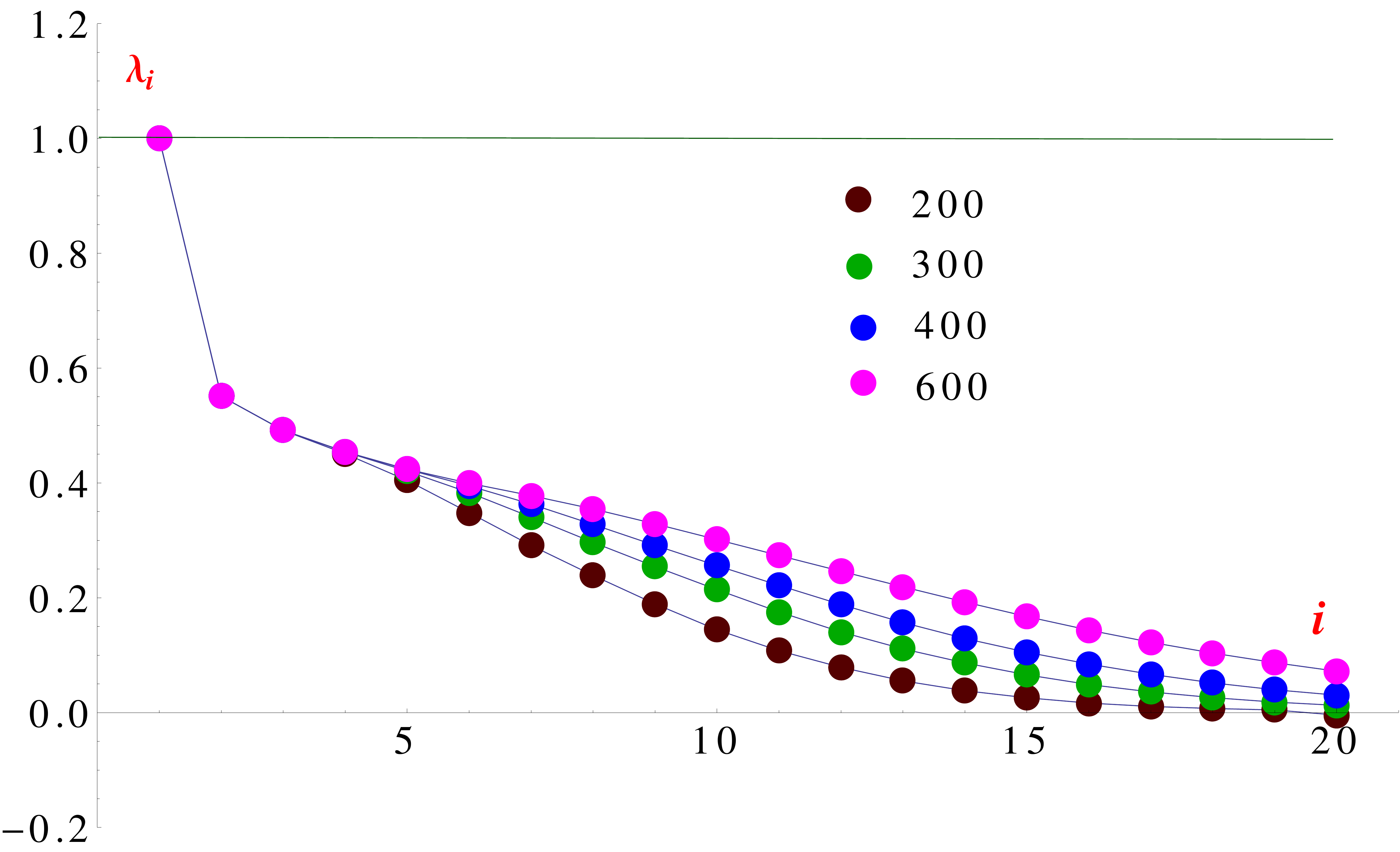}
\end{center}
\caption{Eigenvalues for $d=4$ $m^2=0.05$ and cut-off 
$ n_{max} = 200, ~300, ~400,~600$. The first four eigenvalues are essentially
independent of $n_{max}$.}
\label{eigenfunction-d4}
\end{figure}

We repeat the same analysis for the case $d=4$, $m^2=0.05$.
The dependence of the eigenvalue spectrum at the  critical point 
for various cut-off values $n_{max}$ is shown in Fig.\ \ref{eigenfunction-d4}. 
We see that even at the critical point the first few 
eigenvalues become cut-off independent. This can be attributed 
to a faster fall-off of the transfer matrix elements for large volumes. 
%The dependences of $\lambda_{2}/\lambda_{1}$ and $\lambda_{3}/\lambda_{1}$ 
%on $g$ for this case are presented in Fig. \ref{lambda-d4}. 
%\begin{figure}[h]
%\begin{center}
%\includegraphics[width=0.45\textwidth]{eigen-fitting-d4-005.pdf}
%\end{center}
%\caption{Dependence of the eigenvalue ratios $\lambda_{2}/\lambda_{1}$ and  $\lambda_{3}/\lambda_{1}$ on $g$  for $d=4$ $m^2=0.05$. }
%\label{lambda-d4}
%\end{figure}

\begin{figure}[h]
\begin{center}
\includegraphics[width=0.5\textwidth]{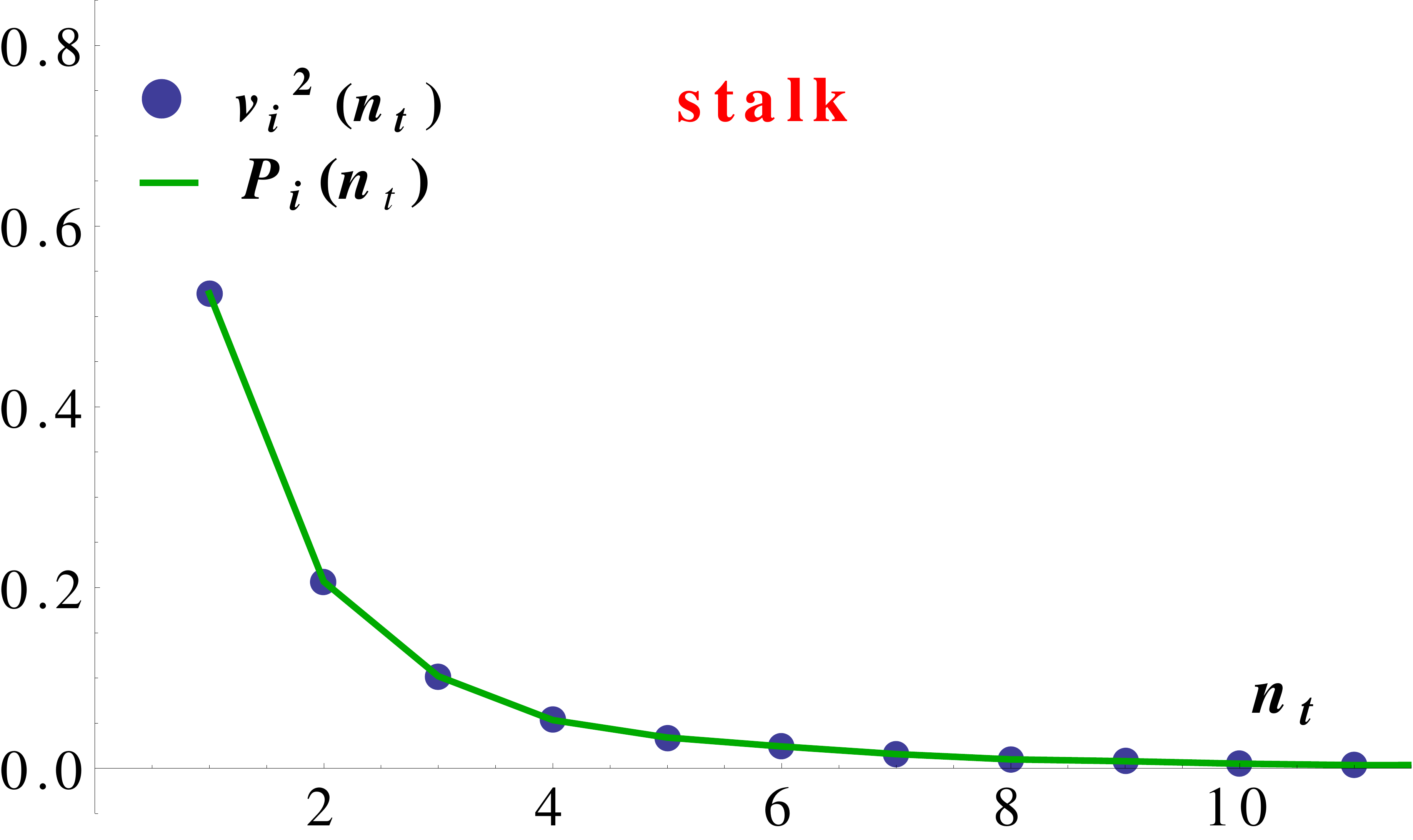}
\end{center}
\caption{The distribution of spatial volume in the ``stalk''  
for $d=4$ and $m^2=0.05$ 
and the square of the first eigenvector of the transfer matrix.}
\label{tail}
\end{figure}

The eigenvalue spectrum is markedly different from 
the pure CDT case since the there is a gap between the 
first and the second eigenvalue even at $g=g_c$, a gap with 
does not vanish for large $n_{max}$. On the other hand the 
rest of the eigenvalues behave like the pure CDT eigenvalues 
in the sense that they coincide for growing $n_{max}$.   
The separation of the  largest eigenvalue  from the rest
is a reflection of the existence of a stalk. In fact
let us denote the first eigenvector (with the largest eigenvalue)
$\nu_1(n)$. Fig.\ \ref{tail} shows that for small $n$ we have 
$\nu_1(n)^2 = P(n)$, where $P(n)$ is the probability distribution 
for spatial volumes {\it in the stalk} (which is almost independent
of $T$ and $N$).   

\section{Conclusions}

In CDT there exists a transfer matrix. This transfer matrix depends
on the geometry of the spatial slices. In 1+1 dimensions the geometry
of the spatial slice is fully characterized by its length (assuming 
the  topology is that of $S^1$). Once we couple the geometry to matter
the transfer matrix still exists, but it will also depend on the 
matter degrees of freedom. Integrating out the matter degrees of 
freedom might introduce non-local interactions and invalidate any 
simple transfer matrix description in terms of geometry only.
However, it turned out that for massive free Gaussian fields 
coupled to geometries, and for the mass not too small there 
is such an {\it effective} transfer matrix which describes very
well the fluctuating geometry of the full model.  

We determined the effective transfer matrix 
in 1+1 dimensional CDT coupled to 4 massive scalar fields
with $m^2 \geq 0.05$. 0.05 was the smallest value of $m^2$ where
we could reliable determine an effective transfer matrix.
We found that the most important term in the effective transfer matrix
was an ``entropic'' factor $1/(n+m)^{\mu}$. $\mu$ is like 
a critical exponent, much like the entropy or susceptibility 
exponent $\gamma$ in non-critical string theory or the 
theory of dynamical triangulations. In the case of non-critical 
string theory $\gamma$ depends on the matter coupled to the 2D 
geometry and there is a phase transition between two completely 
different classes of geometries at $\gamma=0$. For $\gamma > 0$ 
the two-dimensional geometry degenerates into so-called branched polymers.
We have a somewhat similar scenario here: $\mu$ depends on $d$ and $m^2$
and there exists a $\mu_c$ such that for $\mu > \mu_c$ the geometry undergoes
a phase transition and develops a ``blob'' with Hausdorff dimension $d_H=3$.
The appearance of the blob had a profound impact on the 
effective transfer matrix. A gap developed between the two largest 
eigenvalues of the effective transfer matrix and the eigenvector 
corresponding to the largest eigenvalue was essentially equal to the 
square root of the probability distribution of spatial volumes 
of the stalk associated with the blob. We conjecture that a 
similar effective description of the blob--non-blob dynamics will
be present for higher dimensional CDT where it has been shown that 
there also is an effective transfer matrix which describes 
distribution and fluctuation of the spatial volume of the 
time slices \cite{transfer-four}.

\vspace{1cm}

\noindent {\bf Acknowledgments.}
HZ is partly supported by the International PhD Projects
Programme of the Foundation for Polish Science within the European Regional
Development Fund of the European Union, agreement no. MPD/2009/6. JJ ac-
knowledges the support of grant DEC-2012/06/A/ST2/00389 from the National
Science Centre Poland. JA and AG acknowledge support from the ERC-Advance
grant 291092, ``Exploring the Quantum Universe" (EQU)

\end{document}